\documentclass[sigconf]{acmart}
\usepackage{pifont}
\usepackage{xcolor}

\newcommand*{\schedname}{\textsc{RLTune}\@\xspace}
\renewcommand\footnotetextcopyrightpermission[1]{}
\AtBeginDocument{%
  \providecommand\BibTeX{{%
    \normalfont B\kern-0.5em{\scshape i\kern-0.25em b}\kern-0.8em\TeX}}}

\acmYear{2025}\copyrightyear{2025}
\setcopyright{cc}
\setcctype[4.0]{by-sa}
\acmConference[SoCC '25]{ACM Symposium on Cloud Computing}{November 19--21, 2025}{Online, USA}
\acmBooktitle{ACM Symposium on Cloud Computing (SoCC '25), November 19--21, 2025, Online, USA}
\acmDOI{10.1145/3772052.3772257}
\acmISBN{979-8-4007-2276-9/25/11}
\acmConference[SoCC'25]{ACM Symposium on Cloud Computing
}{Nov 19--21, 2025}{Online, USA}
\acmISBN{978-1-4503-XXXX-X/2018/06}

\usepackage{amsmath,amsfonts}
\usepackage{textcomp}
\usepackage{xspace}

\usepackage{graphicx}
\usepackage{tikz}
\tikzset{every node/.append style={font=\sffamily}} 
\renewcommand*\sfdefault{phv}  

\usepackage[font=small,labelfont=bf,textfont=it]{caption}
\usepackage{subcaption}

\usepackage{algorithm}
\usepackage{algpseudocode}

\usepackage{booktabs}
\usepackage{multirow}
\usepackage{makecell}
\usepackage{tabularx}
\usepackage{array}
\usepackage{siunitx}

\usepackage[table]{xcolor}  
\usepackage{soul}           
\usepackage{color}          

\usepackage{url}
\usepackage{balance}
\usepackage[colorinlistoftodos]{todonotes}
\usepackage[utf8]{inputenc}

\usepackage{cleveref}
\crefname{section}{§}{§§}
\Crefname{section}{§}{§§}

\usepackage{comment}  

\usepackage{helvet}      
\begin{document}
\balance
\title{Hybrid Learning and Optimization-Based Dynamic Scheduling for DL Workloads on Heterogeneous GPU Clusters}
\author{Shruti Dongare}
\affiliation{%
  \institution{Virginia Tech, USA}
  \state{}
  \country{}}
  \email{dshruti20@vt.edu}
  
\author{Redwan Ibne Seraj Khan}
\affiliation{%
  \institution{Virginia Tech, USA}
  \state{}
  \country{}}
  \email{redwan@vt.edu}

\author{Hadeel Albahar}
\affiliation{%
  \institution{Kuwait University, Kuwait}
  \state{}
  \country{}}
  \email{hadeel.albahar@ku.edu.kw}
  
\author{Nannan Zhao}
\affiliation{%
  \institution{Northwestern Polytechnical University, China}
  \state{}
  \country{}}
  \email{nannanzhao@nwpu.edu.cn}

  \author{Diego Meléndez-Maita}
\affiliation{%
  \institution{Virginia Tech, USA}
  \state{}
  \country{}}
  \email{dmelendezmaita@vt.edu}
  
\author{Ali R. Butt}
\affiliation{%
 \institution{Virginia Tech, USA}
 \state{}
 \country{}}
 \email{butta@cs.vt.edu}

\begin{abstract}

Modern cloud platforms increasingly host large-scale deep learning (DL) workloads, demanding high-throughput, low-latency GPU scheduling. However, the growing heterogeneity of GPU clusters and limited visibility into application characteristics pose major challenges for existing schedulers, which often rely on offline profiling or application-specific assumptions. We present \schedname, an application-agnostic reinforcement learning (RL)--based scheduling framework that dynamically prioritizes and allocates DL jobs on heterogeneous GPU clusters. \schedname~integrates RL--driven prioritization with MILP--based job-to-node mapping to optimize system-wide objectives such as job completion time (JCT), queueing delay, and resource utilization. Trained on large-scale production traces from Microsoft Philly, Helios, and Alibaba, \schedname improves GPU utilization by up to 20\%, reduces queueing delay by up to 81\%, and shortens JCT by as much as 70\%. Unlike prior approaches, \schedname generalizes across diverse workloads without requiring per-job profiling, making it practical for cloud providers to deploy at scale for more efficient, fair, and sustainable DL\!\ workload management.
\end{abstract}

\ccsdesc[500]{Computing methodologies~Planning and scheduling}
\ccsdesc[500]{Computing methodologies~Machine learning}
\ccsdesc[500]{Computing methodologies~Reinforcement learning}
\ccsdesc[500]{Software and its engineering}

\keywords{Cluster Management, Workload Scheduling, Reinforcement Learning, Deep Learning, Traces}
\maketitle
\section{\textbf{Introduction}}
\label{sec:intro}

Cloud providers increasingly support large-scale deep learning (DL) workloads in shared GPU clusters, powering applications from scientific computing to commercial AI services~\cite{cruz2017accurate,hossain2018environment,rausch2017learning,li2018learning,jurtz2017introduction,aqib2017disaster}. These workloads consume a significant share of data center resources~\cite{datacentertrends,farrell2021mlperf} due to their compute intensity and prolonged runtimes, prompting the development of specialized DL infrastructure~\cite{metasupercomp,googlesupercomp,microsoftsupercomp,teslasupercomp}. 
The extended runtimes and computational intensity of such workloads demand high-performance GPUs for massive parallel processing and memory efficiency~\cite{aach2023large}.

Modern cloud-scale DL clusters often comprise a mix of GPU generations, as operators incrementally upgrade hardware while keeping legacy nodes online to amortize costs and maintain capacity~\cite{Wang2020, Mathuriya2021, Zhang2020, Banerjee2021}. This evolution results in a highly heterogeneous scheduling landscape, where architectural disparities and user preference for faster GPUs~\cite{gao2022deep} create imbalances in resource demand and utilization. In multi-tenant cloud environments, this leads to prolonged queueing delays, degraded performance for certain job classes, and inefficient GPU provisioning at scale. Addressing these challenges is particularly difficult because batch job scheduling is NP-hard, and static policies fail to adapt to the dynamic interplay of job characteristics, cluster state, and workload churn.

Traditional schedulers like Slurm~\cite{slurm} provide scalable, application-agnostic job dispatching and have powered both TOP500 HPC systems~\cite{top500} and commercial cloud backends such as AWS. However, DL workloads introduce new scheduling pressures, e.g., iterative execution~\cite{themis,optimus}, gang scheduling~\cite{gang_scheduling}, intra-node GPU fragmentation~\cite{chaudhary2020balancing,antman}, and resource sharing~\cite{Alibaba23} that demand more adaptive and workload-aware mechanisms. These new characteristics has led to DL-focused schedulers such as Gavel~\cite{gavel} and Sia~\cite{sia}, which use predictive, preemptive strategies to optimize throughput and fairness. Yet, these systems rely on profiling, matching new jobs to previously seen ones based on model architecture, dataset, or runtime behavior. In practice, cloud platforms lack detailed metadata due to user privacy, model diversity, and rapidly changing workloads, as seen in real-world traces like Microsoft Philly~\cite{philly} and Alibaba PAI~\cite{MLaaS}. As a result, profiling-dependent methods limit scalability, and yield unreliable approximations. Many DL jobs are black boxes (e.g., proprietary models or novel architectures), making profiling infeasible and motivating learning-based alternatives that operate effectively in production clouds. While profiling already faces limitations in diverse clusters, its relevance further reduces as emerging deployment trends increasingly dedicate separate pools to specialized workloads such as LLM serving or DLRM training to improve stability. This specialization narrows cross-application diversity, leaving fewer opportunities for profiling-based schedulers to provide benefit. 
In contrast, dynamic, profiling-free schedulers that adapt to runtime signals remain effective across both mixed and dedicated environments.

\begin{table*}[!ht]
\centering
\caption{Comparison of Existing ML/DL Job Schedulers and RL-Based CPU Schedulers with \schedname. }
\vspace{-0.7em}
\begin{tabularx}{\textwidth}{|l|X|c|c|c|c|c|c|}
\hline
\textbf{Scheduler} & \textbf{Scheduling Strategy} & \textbf{GPU Het.} & \textbf{App-Agn.} & \textbf{Offline Prof. free} & \textbf{Dyn. Policy} & \textbf{Preempt.} & \textbf{Elastic} \\
\hline
Slurm           & Multi-factor       & \ding{51} & \ding{51} & \ding{51} & limited & configurable & \ding{55} \\
QSSF\cite{helios}            & Historic data      & \ding{51} & \ding{51} & \ding{55} & \ding{55} & \ding{55} & \ding{55} \\
Gavel\cite{gavel}           & Gavel              & \ding{51} & \ding{55} & \ding{55} & \ding{51} & \ding{51} & \ding{55} \\
Pollux\cite{pollux}         & Pollux             & \ding{55} & \ding{55} & \ding{55} & \ding{51} & \ding{55} & \ding{51} \\
Sia\cite{sia}               & Sia                & \ding{51} & \ding{55} & \ding{55} & \ding{51} & \ding{51} & \ding{51} \\
\textbf{\schedname (Ours)}  & RL + MILP          & \ding{51} & \ding{51} & \ding{51} & \ding{51} & \ding{55} & \ding{55} \\
SchedInspector\cite{schedinspector}  & RL        & \ding{55} (CPU) & \ding{51} & \ding{51} & limited  & \ding{55} & \ding{55} \\
RLScheduler\cite{RLScheduler}        & RL        & \ding{55} (CPU) & \ding{51} & \ding{51} & limited & \ding{55} & \ding{55} \\
\hline
\end{tabularx}
\label{tab:scheduler-comparison}
\vspace{-0.6em}
\end{table*}

Although Gavel and Sia rely on performance prediction, \schedname instead follows an application-agnostic design, in the DL scheduling domain, which means that their decisions do not depend on model semantics such as architecture, dataset, optimizer, batch size, or training objective. \schedname is also profiling-free as it does not build or depend on per-job performance models, instead using only user-submitted metadata and queue-level information to guide scheduling in real time. Table~\ref{tab:scheduler-comparison} summarizes the design trade-offs across representative DL schedulers. Application-agnostic methods scale to real-world traces lacking detailed metadata, but require more intelligent mechanisms to adaptively prioritize jobs and make fine-grained allocation decisions in heterogeneous, resource-constrained clusters.

We argue for a complementary learning-driven yet application-agnostic approach to DL job scheduling. Reinforcement learning (RL)\cite{RF} is well suited for this goal. It learns from experience rather than static heuristics, operates in partially observable environments, and optimizes long-term outcomes like JCT and resource utilization. Prior works demonstrated the effectiveness of RL in CPU scheduling (e.g., RLScheduler~\cite{RLScheduler}, SchedInspector~\cite{schedinspector}), applying RL to GPU scheduling introduces new challenges such as co-allocation constraints, heterogeneity, and fragmentation arising from the hierarchical, multi-node nature of GPU clusters. These factors fundamentally change the learning problem and reshape the reward dynamics, thus requiring novel design. To address these, we present \schedname, a dynamic scheduler that couples RL-based dynamic prioritization (DP) with mixed-integer linear programming (MILP)–based resource allocation. \schedname leverages job-level and system-level signals (e.g., user metadata, resource availability, queue state) to construct engineered features and select the most relevant ones as input to the RL agent for Dynamic Prioritization (DP), while MILP performs multi-dimensional allocation across GPUs, CPUs, and memory in alignment with cluster-level objectives. While RL and MILP have been individually applied to scheduling, \schedname uniquely separates and couples them using RL for proactive prioritization and MILP for multi-dimensional look-ahead allocation forming a unified hybrid framework that learns without per-application profiling and generalizes across heterogeneous clusters.

To evaluate \schedname, we use three diverse, publicly available DL workload traces: Philly~\cite{philly}, Helios~\cite{helios}, and Alibaba'20~\cite{MLaaS}, and assess performance using metrics waiting time, JCT, bounded slowdown (BSLD), and resource utilization. Our evaluation examines how \schedname captures fine-grained opportunities for long-term efficiency through safe, reward-driven trade-offs in job prioritization and allocation. We compare the trained RL policy against state-of-the-art baselines~\cite{slurm, helios, RLScheduler, schedinspector}, demonstrating improvements across multiple performance objectives.

Specifically, \schedname makes the following contributions:

\begin{itemize}
    \item It captures key scheduling challenges in heterogeneous GPU clusters such as application diversity, resource fragmentation, and profiling infeasibility, motivating an application-agnostic learning approach.
    \item Couples RL-based dynamic prioritization with MILP-based multi-resource allocation, leveraging runtime features for long-term optimization without per-application profiling.
    
    \item We train and evaluate \schedname on real-world traces, measuring queueing delay, JCT, BSLD, and resource utilization.
    \item \schedname achieves up to 81\% lower queueing delay, 70\% shorter JCT, and 20\% higher GPU utilization than state-of-the-art schedulers, across diverse workloads and cluster setups.
    \item We deploy \schedname on a heterogeneous Slurm cluster, demonstrating its effectiveness in improving end-to-end scheduling under real-world conditions.

\end{itemize}

\section{\textbf{Background and Motivation}}
\label{sec:motivation}
\begin{figure}[!ht]
    \centering
    \begin{minipage}{0.5\textwidth}
        \centering
        \includegraphics[width=\linewidth]{figures/new_motivation/spread1.pdf}
        \vspace{-1em}
        \caption{Slurm’s Multi-Factor Priority scheduling (per-job-per-node allocation)}
        \label{fig:spread1}
    \end{minipage}
    \hfill
    \begin{minipage}{0.5\textwidth}
        \centering
        \includegraphics[width=\linewidth]{figures/new_motivation/spread2.pdf}
        \vspace{-1em}
        \caption{Jobs scheduled with Priority shuffling  (per-job-per-node allocation)}
        
        \label{fig:spread2}        
    \end{minipage}
    \vspace{-0.5em}
\end{figure}
\noindent\textbf{Empirical Gaps in De Facto Production Scheduling.} In this section, we examine current production scheduling to identify remaining improvement opportunities, focusing on Slurm which is the de facto foundation for modern GPU clusters. While prior schedulers employ predictive or profiling-based strategies, we rethink how production systems can operate under profiling-free, dynamically evolving conditions. To this end, we empirically analyze where even this mature baseline leaves room for adaptive improvement through controlled experiments on a Slurm-based cluster. Our goal is to observe how its multifactor priority plugin and allocation mechanisms perform under realistic DL workloads. We deployed Slurm (21.08.5) on two P100 nodes (4 GPUs each), configured with \textit{SchedulerType=sched/backfill}, \textit{PriorityType=priority/multifactor}, and \textit{SelectType=select/cons\_tres}. We submitted a mix of DL fine-tuning and inference workloads, including LLM jobs requesting single-GPU, multi-GPU, and multi-node execution.

\begin{figure}[!ht]
    \centering
    \begin{minipage}{0.5\textwidth}
        \centering
        \includegraphics[width=\linewidth]{figures/new_motivation/pack1.pdf}
        \vspace{-1em}
        \caption{Slurm’s Multi-Factor Priority scheduling (packing allocation)}
        \label{fig:pack1}
    \end{minipage}
    \hfill
    \begin{minipage}{0.5\textwidth}
        \centering
        \includegraphics[width=\linewidth]{figures/new_motivation/pack2_correct.pdf}
        \vspace{-1em}
        \caption{Jobs scheduled with Priority shuffling (packing allocation)}
        \label{fig:pack2}        
    \end{minipage}
\end{figure}

Fig.~\ref{fig:spread1} and Fig.~\ref{fig:pack1} illustrate Slurm’s scheduling behavior under two configurations. In Fig.~\ref{fig:spread1}, Slurm operated in its default per-job-per-node mode with \textit{OverSubscribe=No}, which prevents multiple jobs from sharing a node and causes jobs to be distributed across separate nodes, leaving several GPUs idle. As a result, resource utilization decreased and cumulative waiting time rose sharply. When \textit{OverSubscribe=Yes} was enabled (Fig.~\ref{fig:pack1}), analogous to a gang-scheduling scenario, Slurm packed multiple jobs per node, improving GPU utilization but introducing intra-node contention that could degrade performance. These observations highlight the limited flexibility of existing policies in balancing the classical spread–versus–pack trade-off.
We next examined job prioritization, determined in Slurm by its multi-factor priority plugin. To test whether small priority adjustments can yield benefits, we smartly altered job priorities. In the per-job-per-GPU case (Fig.~\ref{fig:spread2}), swapping the priorities of Jobs 327 and 328, and in the packed case (Fig.~\ref{fig:pack2}), raising Job 349 above 348, both led to noticeable reductions in cumulative waiting time and makespan. The dotted red lines (JCT) and purple regions (waiting time) show that even minor, context-driven priority changes can produce measurable gains.

These experiments reveal two key gaps in current scheduler behavior: (1) the absence of dynamic, context-aware priority adjustment, and (2) limited flexibility in deciding when to isolate or pack jobs. While Slurm’s heuristics are robust and general-purpose, they lack the adaptive intelligence to exploit small but high-impact scheduling opportunities, especially at scale. Recognizing and addressing these gaps at runtime could yield substantial performance gains. Our findings highlight opportunities to introduce context-aware control into an otherwise mature field of GPU scheduling, enabling systems like Slurm to adapt intelligently using runtime feedback instead of per-job profiles. This naturally raises the question of what mechanism can learn and react to system feedback quickly enough to guide scheduling decisions in real time.

\noindent\textbf{Investigating Learning-Driven Scheduling.}  Building on our observation in prior subsection, we next explore whether learning-based mechanisms can enable the level of adaptivity required for profiling-free GPU scheduling.
RL is a natural candidate because it can learn directly from runtime feedback and optimize long-term performance.
RL has already proved successful in CPU scheduling \cite{RLScheduler, schedinspector, DRAS}. This prior success reveals an opportunity, however, GPU clusters present fundamentally different scheduling challenges, making it non-trivial to apply existing RL policies directly. 

\begin{table}[ht]
\centering
\footnotesize
\setlength{\tabcolsep}{3pt}
\renewcommand{\arraystretch}{1.1}
\caption{Summary of CPU and GPU traces comparing wait time, run time, job arrival rate, and total allocated resources (CPUs or GPUs).}
\begin{tabular}{lccccccc}
\toprule
\textbf{Trace} & \textbf{Type} & \textbf{Jobs} & \textbf{Avg Wait} & \textbf{Avg Run} & \textbf{Arrival Rate} & \textbf{Total Alloc} \\
 & & & (s) & (s) & (jobs/s) & (CPUs/GPUs) \\
\midrule
SDSC-SP2   & CPU & 2887    & 10403.6 & 7319.6  & 0.001123  & 44416 \\
HPC2N      & CPU & 1768    & 13985.4 & 8699.6  & 0.000868  & 38403 \\
Philly     & GPU & 60k   & 2703.3  & 26299.2 & 0.022333  & 84602 \\
Helios     & GPU & 85k   & 99.3     & 2481.4   & 0.032919  & 139996 \\
Alibaba    & GPU & 200k  & 22.1    & 5466.3  & 0.077136   & 214372 \\
\bottomrule
\end{tabular}
\label{tab:cpu_gpu}
\end{table}

\begin{figure}[!ht]
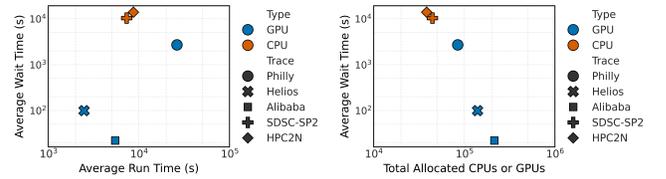

    \centering
    \begin{subfigure}[t]{0.49\linewidth}
        \centering
        \includegraphics[width=\linewidth]{figures/new_motivation/cpu_gpu_a.pdf}
        \caption{Average run time vs.\ average wait time.}
        \label{fig:cpu_gpu_a}
    \end{subfigure}
    \hfill
    \begin{subfigure}[t]{0.49\linewidth}
        \centering
        \includegraphics[width=\linewidth]{figures/new_motivation/cpu_gpu_b.pdf}
        \caption{Allocated resources vs. \ wait time.}
        \label{fig:cpu_gpu_b}
    \end{subfigure}
    
    \vspace{-0.6em}
    \caption{Comparison of CPU and GPU workload traces}
    \label{fig:cpu_gpu}
    \vspace{-1em}
\end{figure}

To empirically validate these differences, we analyzed two CPU traces (SDSC-SP2, HPC2N)~\cite{cpu_traces} and three GPU traces (Philly, Helios, Alibaba). Table~\ref{tab:cpu_gpu} reports average wait time, run time, arrival rate, and total allocated resources, normalized over one month. Fig.~\ref{fig:cpu_gpu} provides a complementary view, comparing (a) run time vs. wait time and (b) allocated resources vs. wait time.
The results reveal distinct scheduling dynamics: GPU traces exhibit much higher job counts, faster arrivals, and greater aggregate resource demand. The run–wait relationship also diverges where CPU jobs experience long waits even for short runtimes, whereas GPU workloads vary by cluster (Philly: long runs with moderate waits; Helios and Alibaba: short runs with minimal waiting). These patterns highlight different queuing and contention behaviors in multi-GPU, multi-tenant environments.
Our experiments further confirmed that CPU-trained RL models failed to converge on heterogeneous GPU clusters. This shift fundamentally redefines the learning problem by changing what RL predicts, how rewards evolve, and which signals matter.
Resource needs in CPU scheduling are uniform, however, in GPU clusters, scheduling extends to multi-dimensional allocation across GPUs, CPUs, memory, and interconnects, as well as the spread–versus–pack trade-off.
To address this, \schedname employs RL for dynamic prioritization while delegating multi-resource allocation to a complementary solver. Even with this reformulation, a key question remains: can it remain effective under the bursty, non-stationary conditions of real GPU workloads?

\noindent\textbf{Workload Variations and Scheduling Stability.}
\label{sec:waiting_time_patterns}
\begin{figure}[!ht]
	\centering
	\includegraphics[width=0.5\textwidth]{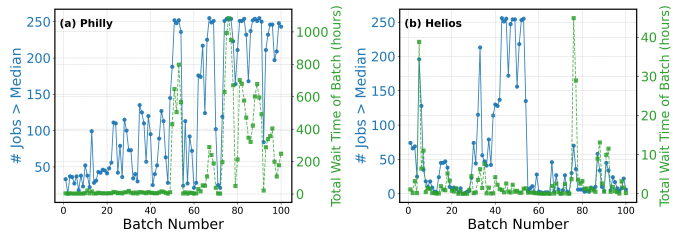} 
  \vspace{-1em}
	\caption{Batch-wise Analysis: Jobs > Median Wait vs Total Wait Time}
	\label{fig:wait_time_moti}
\end{figure}
The effectiveness of any scheduler depends on how well it adapts to evolving job arrivals and queue dynamics. To examine this, we analyzed scheduling trajectories, i.e., the temporal sequences of job arrivals, queue responses, and cumulative waiting times, from two large GPU traces, Philly and Helios. For 100 consecutive batches of 256 jobs each, we measured (1) the number of jobs per batch exceeding the global median wait time and (2) the total cumulative wait within the same window.
As shown in Fig.~\ref{fig:wait_time_moti}, these trajectories vary sharply: some batches remain nearly flat with few jobs waiting and minimal cumulative wait time, while others exhibit severe congestion where most jobs wait, accumulating hundreds of hours of total delay. Such non-stationary behavior reveals that workload pressure in GPU clusters is highly bursty and unpredictable. Therefore, Workload characterization alone is insufficient. 

Such variability directly affects a learning-based scheduler’s ability to generalize. A policy that performs well during steady phases may fail under sudden congestion or resource fragmentation. For instance, if a scheduler encounters consecutive trajectories where most jobs experience low waiting times, it may appear effective even when its decisions have little real impact and overfitting to easy scenarios and underperforming once the system becomes bursty or imbalanced.
Reinforcement learning, however, can leverage this variability when guided by timely feedback to learn stable trade-offs across changing workload conditions. These insights motivate our subsequent design, where we encode workload dynamics into the reward formulation to sustain performance under bursty, dynamic, and unpredictable environments.

\noindent\textbf{Toward a Feasible Hybrid Scheduling Framework.} To translate insights from the preceding analysis into practice, we must verify whether dynamic prioritization and adaptive allocation remain feasible at scale under real system constraints.
In real deployments, GPU scheduling becomes a multi-dimensional decision problem: when multiple jobs are colocated on a node, GPUs alone are not the bottleneck but CPU, memory, and intra-node bandwidth can introduce interference and affect runtime. In our experiments, Fig.~\ref{fig:pack1} and Fig.~\ref{fig:pack2}, we adopt a GPU proportionate CPU allocation strategy and a relaxed memory policy, allowing dynamic use up to a safe threshold. While adequate for feasibility testing, this approach reveals the need for fine-grained modeling of CPU and memory coupling in job-to-GPU mapping. Solver-based, look-ahead allocation can address this challenge by anticipating contention and jointly optimizing GPU, CPU, and memory placements. This motivates extending the scheduling formulation through a Mixed-Integer Linear Programming (MILP)-based allocation framework.

Given this complexity, reinforcement learning remains well suited to control the priority function, dynamically adjusting to workload and system feedback.
However, RL alone cannot guarantee global resource-level optimality across dimensions.
We therefore adopt a hybrid strategy that couples RL-based dynamic prioritization with MILP-based multi-resource allocation.
This separation allows RL’s adaptability to complement MILP’s stability, improving interpretability, reducing RL training complexity, and ensuring consistent performance across heterogeneous clusters.
These insights motivate \schedname’s design, which we describe in the next section.
\section{\textbf{Design}}
\label{sec:design}



\subsection{\schedname System Overview and Job Lifecycle}


\begin{figure}[!ht]
	\centering
	\includegraphics[width=0.5\textwidth]{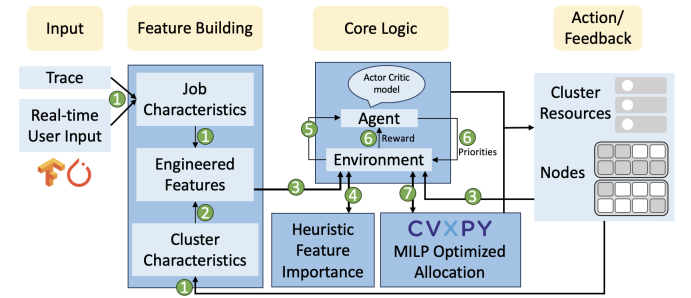} 
    \vspace{-1em}
	\caption{System Overview and life cycle of Job in
\schedname}
	\label{fig:System Overview}
\end{figure}

We integrate RL-based prioritization with resource-aware placement in \schedname, a hybrid learning-and-optimization scheduler designed for heterogeneous GPU clusters. Fig.~\ref{fig:System Overview} presents an overview of \schedname, highlighting its components and the job life-cycle under its operation. The system is organized into three key modules: Feature Building, Core Logic (RL+MILP), and Action/Feedback Network. The Feature Building module extracts job- and cluster-level characteristics, while Feature Sampling dynamically selects the most informative features at runtime based on the current cluster state.
The RL agent, implemented using an Actor–Critic architecture with PPO, assigns job priorities adaptively according to real-time cluster conditions.
The Allocation Optimization module, formulated as a Mixed-Integer Linear Program (MILP), then determines the most efficient job-to-node mapping given available resources. \schedname operates in two phases: training and evaluation. We describe the step-by-step workflow of both phases and the contribution of each system component in the following sections.
We use (\tikz[baseline=(char.base)]{
\node[shape=circle,draw,inner sep=1pt] (char) {1};}) to denote steps corresponding to those in Fig.~\ref{fig:System Overview}.

\begin{figure}[!t]
	\centering
	\includegraphics[width=0.49\textwidth]{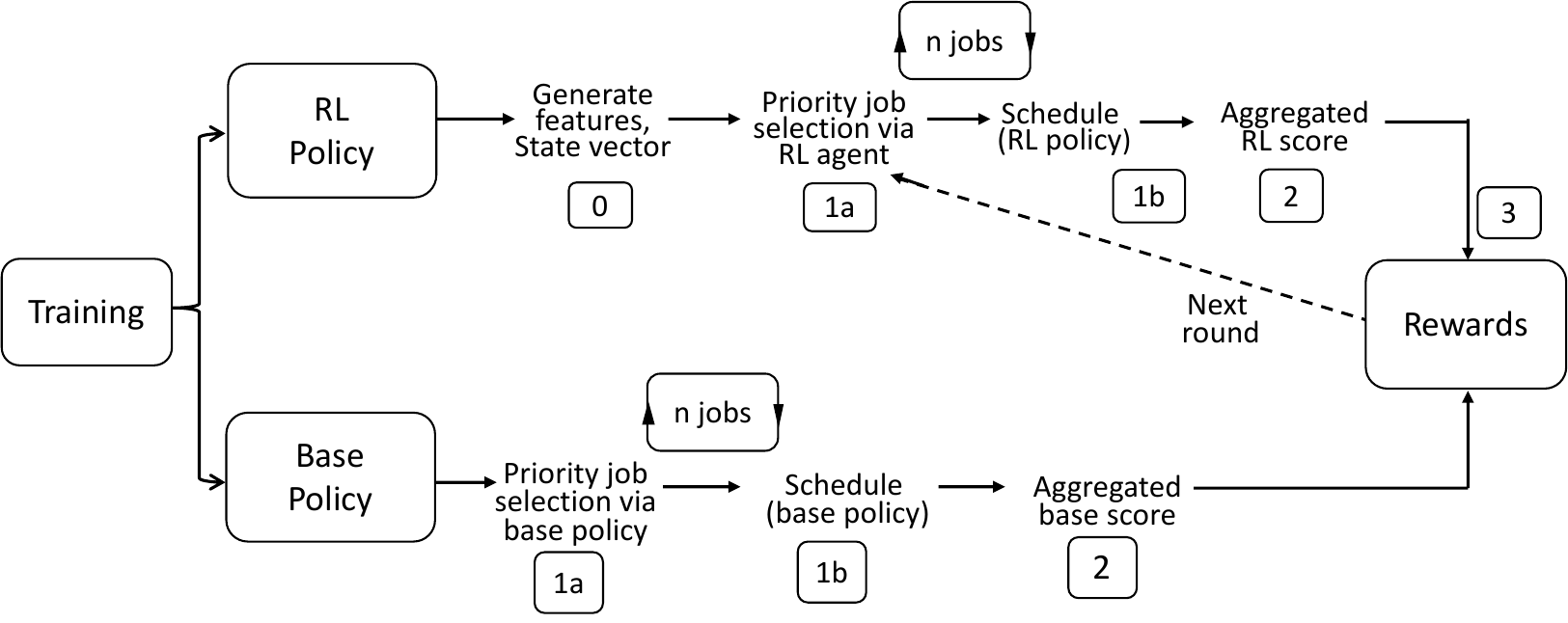} 
  \vspace{-1.5em}
	\caption{Training pipeline a in simulated environment for one batch}
    \vspace{-0.7em}
	\label{fig:execution}
\end{figure}


\subsubsection{\textbf{Training Phase and Workflow}}
\label{subsec:training}
Training an RL agent requires extensive iterations and large volumes of data; hence, we conduct training in an RL environment that mimics a Slurm simulator using large trace inputs.
We allocate 90\% of the trace data for training and perform periodic evaluations after every \textit{N} batches (\textit{N} ranges from 100–1000).
The remaining 10\% of the trace data is held as unseen data for evaluation. 

Fig.~\ref{fig:execution} illustrates the step-by-step workflow of how jobs traverse each component during training. We train the RL agent in batches of 256 jobs, with one epoch consisting of 100 such batches. 
Each batch is processed through two pipelines: the base policy pipeline and the RL policy pipeline. Following an application-agnostic approach, the Feature Building Module (FBM) (see Fig.\ref{fig:System Overview}) in both pipelines scans only visible job features such as job ID, submit time, requested resources, etc.(\tikz[baseline=(char.base)]{
\node[shape=circle,draw,inner sep=1pt] (char) {1};}). Simultaneously, the FBM scans the cluster state to extract system-level characteristics (\tikz[baseline=(char.base)]{
\node[shape=circle,draw,inner sep=1pt] (char) {1};}). In Fig.~\ref{fig:execution}, step \tikz[baseline=(char.base)]{ \node[shape=rectangle,draw,inner sep=1pt] (char) {0};} of the RL pipeline combines these scanned features to construct comprehensive engineered features for each incoming job (\tikz[baseline=(char.base)]{
\node[shape=circle,draw,inner sep=1pt] (char) {2};}), explained more in \cref{subsec:sys_comp}. Core logic component handles steps \tikz[baseline=(char.base)]{ \node[shape=rectangle,draw,inner sep=1pt] (char) {1a};}, \tikz[baseline=(char.base)]{ \node[shape=rectangle,draw,inner sep=1pt] (char) {1b};} of RL pipeline which differ from \tikz[baseline=(char.base)]{ \node[shape=rectangle,draw,inner sep=1pt] (char) {1a};}, \tikz[baseline=(char.base)]{ \node[shape=rectangle,draw,inner sep=1pt] (char) {1b};} of base pipeline.
The core logic includes the RL environment, RL agent and feature sampling module. RL environment continuously monitor arrived jobs and adds them to \textit{job\_queue} of both pipelines. The base pipeline uses one of several scheduling policies (e.g., FCFS, SJF, WFP, UNI-CEP, F1, QSSF). Each policy’s priority function is derived from one or more job features (e.g., Submit Time (ST), Requested or Run Time (RT), Wait Time (WT), Requested Resources (N)). During execution of \tikz[baseline=(char.base)]{ \node[shape=rectangle,draw,inner sep=1pt] (char) {1a};}, \tikz[baseline=(char.base)]{ \node[shape=rectangle,draw,inner sep=1pt] (char) {1b};}, top priority job from \textit{job\_queue} is selected for scheduling, depending on the selected base policy's priority function. In the RL pipeline, the environment takes the feature set from the FBM, applies feature sampling to select a fixed number of important features for each job, and aggregates them across all jobs in the \textit{job\_queue} to form the state matrix $S_t$ (\tikz[baseline=(char.base)]{ \node[shape=circle,draw,inner sep=1pt] (char) {3};}),(\tikz[baseline=(char.base)]{ \node[shape=circle,draw,inner sep=1pt] (char) {4};}); details are in \ref{subsec:sys_comp}. The state matrix $S_t$ is forwarded to the RL agent as input (\tikz[baseline=(char.base)]{ \node[shape=circle,draw,inner sep=1pt] (char) {5};}).
The agent outputs a corresponding priority vector $A_t$  (\tikz[baseline=(char.base)]{ \node[shape=circle,draw,inner sep=1pt] (char) {6};}), ranking jobs in real time. The top-K jobs from this ranked list are forwarded to the MILP-based Allocation Optimizer, which evaluates multi-dimensional, look-ahead strategies to select the optimal job-to-node mapping under current GPU, CPU, and memory constraints(\tikz[baseline=(char.base)]{ \node[shape=circle,draw,inner sep=1pt] (char) {7};}).
The simulator tracks resource availability, moves jobs from \textit{job\_queue} to \textit{running\_queue}, and maintains allocation and release records. 
Upon successful scheduling, a score (base or RL) is computed using a chosen performance metric such as wait time, completion time, bounded slowdown, or resource utilization representing the base score in the base pipeline and the RL score in the RL pipeline.
This completes the execution of step \tikz[baseline=(char.base)]{ \node[shape=rectangle,draw,inner sep=1pt] (char) {1b};} in both pipelines. In step \tikz[baseline=(char.base)]{ \node[shape=rectangle,draw,inner sep=1pt] (char) {2};} of both base and RL pipelines, individual job scores are aggregated to compute the Aggregated Base Score (ABS) and the Aggregated RL Score (ARS) separately. Next, we calculate reward by subtracting ARS from ABS \tikz[baseline=(char.base)]{ \node[shape=rectangle,draw,inner sep=1pt] (char) {3};}. Details of the reward function are provided in \cref{subsec:sys_comp}. During the next iteration in the RL pipeline, the rewards are fed back to reinforce or adjust the agent’s actions(\tikz[baseline=(char.base)]{ \node[shape=circle,draw,inner sep=1pt] (char) {6};}). Each batch completes a full feedback loop, and an epoch consists of 100 such batches. Training typically spans epochs until the policy converges.

\subsubsection{\textbf{Evaluation Phase and Workflow}}
\label{subsec:eval}
During the evaluation phase, the base and RL pipelines run independently to compare performance between the baseline policy and \schedname{}. Evaluations are conducted either in simulation or on a live Slurm deployment. In RL pipeline, steps \tikz[baseline=(char.base)]{ \node[shape=rectangle,draw,inner sep=1pt] (char) {0};},\tikz[baseline=(char.base)]{ \node[shape=rectangle,draw,inner sep=1pt] (char) {1a};},\tikz[baseline=(char.base)]{ \node[shape=rectangle,draw,inner sep=1pt] (char) {1b};},\tikz[baseline=(char.base)]{ \node[shape=rectangle,draw,inner sep=1pt] (char) {2};} mirror those of the training phase, reusing the same Feature Building and Feature Sampling modules to construct the state matrix $S_t$. The RL agent then produces a ranked list of job priorities based on $S_t$, and the top-K prioritized jobs are forwarded to the MILP optimizer for job-to-node placement under current cluster constraints. 
In real-time Slurm evaluation, the job queue is scanned every minute to generate $S_t$, capturing both waiting and newly arrived jobs. The RL agent updates job priorities, which are applied directly to Slurm using the \textit{scontrol --priority=} command and  allocation flexibility is managed through Slurm’s \textit{--oversubscribe} flag as per the solver’s guidance. This ensures that updated priority and allocation decisions are refreshed for submitted jobs before they start running, avoiding any data movement across GPUs, nodes, or storage systems. During evaluation, the aggregated score computed as the sum of individual job score (e.g., per-job wait time) directly represents batch-level performance (e.g., total or average wait time). The batch size can be tuned based on the workload arrival rate. \schedname also handles SLA-bound or high-priority jobs that cannot tolerate its operational overhead through the baseline scheduler, ensuring fairness and compliance.

\subsection{System Components}
\label{subsec:sys_comp}


\paragraph{\textbf{Feature Building and Feature Sampling for Importance}}
To help the RL agent make effective scheduling decisions, we design a feature-building module that captures both job and cluster characteristics. This structured feature set enables the agent to learn key patterns and dependencies in the scheduling environment. For every incoming job, the module constructs a set of engineered features from runtime attributes, resource demands, and cluster status indicators that collectively describe the current scheduling context. The complete list of features is summarized in Table~\ref{tab:features_table}.

\begin{table}[ht]
    \centering
    \caption{Feature Categories and Corresponding Features.}
    \vspace{-0.6em}
    \label{tab:features_table}
    \begin{tabular}{l p{0.55\linewidth}}
        \toprule
        \textbf{Feature Category} & \textbf{Features} \\
        \midrule
        Visible Job Features      & job ID, user info, requested GPUs, virtual cluster, GPU type, requested time, submit time, req\_CPU, req\_mem. \\
        Cluster Characteristics   & free nodes, can\_schedule\_now, num\_ways\_to\_schedule. \\
        Engineered Features       & Demand-Supply Ratio (DSR), Job Size, Job Urgency Score, Future Availability, Cluster Fragmentation Factor (CFF). \\
        \bottomrule
    \end{tabular}
    \vspace{-1em}
\end{table}

The primary features such as requested GPUs, submit time, and job duration are obtained directly from trace data during training or scanned from the job in real time. Cluster features are extracted from the current system state, including the number of free and used GPUs or nodes. Computed features like can\_schedule\_now and num\_ways\_to\_schedule capture job feasibility under current resource constraints. From primary features, we derive engineered features by combining or transforming base features to express more actionable scheduling insights. We first describe three complex engineered features through equations, followed by the remaining ones. The first engineered feature, Demand–Supply Ratio (DSR), is defined as follows:

{\small
\[
\text{demand\_supply\_ratio} = \left( \frac{\text{[req. gpus]}_{\text{type}}}{\text{[free gpus]}_{\text{type}}} \right)_{\text{norm}} \quad (1)
\]
}
\vspace{-0.1\baselineskip}
Demand Supply Ratio captures scaled measure of the relationship between the demand for GPUs of a specific type and the current availability of GPUs of the same type. This feature provides insight for keeping balance between cluster load and resource contention. Second feature is Future Availability defined as follow:
{\small
\[
\text{future\_avail}=
\left[
\sum_{\text{type}}
\Big(
\text{free\_gpus}
- (j_{\text{curr}})_{\text{req\_gpus}}
- \sum_{j \in \text{nodes}} (j_i)_{\text{req\_gpus}}
\Big)
\right]_{\text{norm}}\!(2)
\]
}
\vspace{-0.1\baselineskip}
The Future Availability estimates the expected number of free GPUs in the cluster after accounting for the GPUs requested by the current job and those already allocated to other jobs on the same nodes. By incorporating this forward-looking estimate, the RL agent gains a predictive view of resource usage if the current job is scheduled. Next Cluster Fragmentation Factor (CFF) defined as follow:
{\small
\[
\text{CFF} = \left( 1 - \frac{\sum_{\text{nodes}} (\text{free\_gpus})^2}{\text{total\_free\_gpus}} \right)_{\text{norm}} \quad (3)
\vspace{1\baselineskip}
\]
}
CFF measures how evenly free GPUs are distributed across the cluster. A higher CFF indicates greater fragmentation, where GPUs are scattered across multiple nodes, making it harder to schedule large multi-GPU jobs on the same node. 


Not all features listed in Table~\ref{tab:features_table} are directly used for state construction. In total, we maintain 17 features across the system: some are used purely for metadata tracking (e.g., Job Id, user info, virtual cluster), and such as Req\_CPU and Req\_mem are used only when explicitly provided, otherwise inferred from GPU share. For forming the state vector input to the RL agent, we employ heuristic-based feature sampling to select a subset of eight key features from the complete set. This selection avoids redundancy by choosing either raw or engineered variants based on situational relevance. For instance, when CFF is high, the sampler select and weights feature \texttt{Job\_size = normalized(requested\_gpus $\times$ requested\_time)} to favor short jobs that can efficiently fill fragmented nodes. Under low-fragmentation conditions, the job urgency score is emphasized to boost the priority of aged jobs while avoiding penalties for large or long-running jobs. When a job can be scheduled in multiple valid ways, importance is assigned to the num\_ways\_to\_schedule feature. This increases the likelihood that such flexible jobs receive higher priority, allowing the scheduler to exploit placement opportunities and improve overall cluster utilization.

We selected 8 features as a balanced trade-off between model simplicity and expressive capacity. A preliminary sensitivity analysis comparing feature sets of sizes 7, 8, and 9 showed that 8 features consistently delivered stable and efficient learning, with no significant benefit from including more. This compact yet expressive representation reduces runtime overhead while preserving enough information to support robust policy learning.

\paragraph{\textbf{Actor Critic}}

\begin{figure}[!ht]
	\centering
	\includegraphics[width=0.48\textwidth]{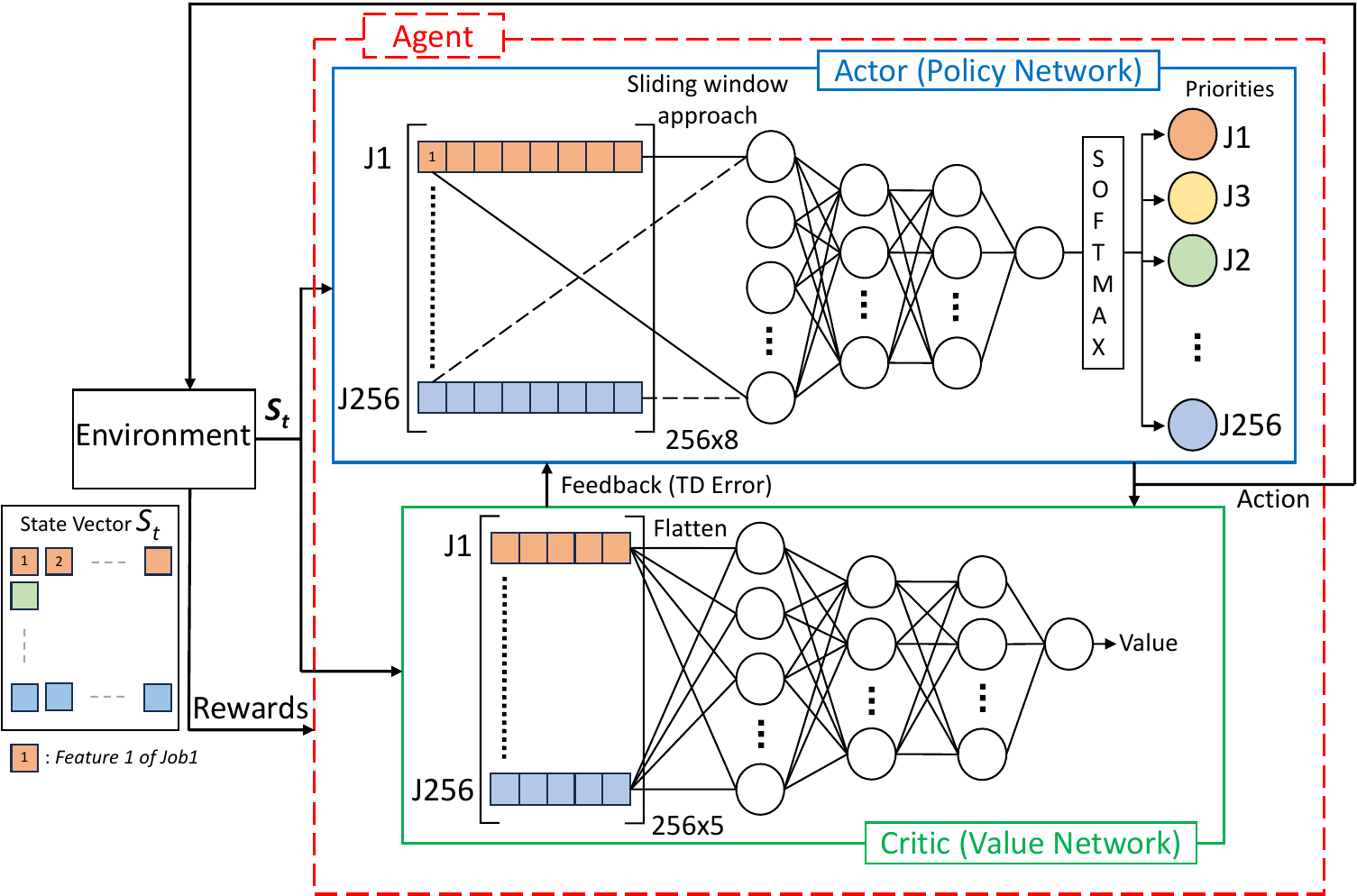} 
  \vspace{-1em}
	\caption{Actor-Critic Architecture}
	\label{fig:Actor-Critic}
 \vspace{-0.3cm}
\end{figure}

The goal of \schedname is to enable dynamic prioritization by learning and adapting scheduling decisions to changing cluster availability. To achieve this, we employ the Proximal Policy Optimization (PPO) algorithm~\cite{ppo} with an Actor–Critic framework. The RL agent, modeled as an Actor–Critic network, consists of an actor that assigns priorities to queued jobs and a critic that evaluates the quality of these decisions to guide policy updates.

The environment provides a State matrix $S_t$, composed of a feature vector that we then split into Observation vector (OV) of fixed size (\(batch size \times 8 job features\)), and a critic vector (CV) of size (\(batch size \times 5 job features\)). The OV contains normalized values of the eight key features selected through heuristic-based sampling. The CV includes five core features such as submit time, run time, and can\_schedule\_now. CV estimate long-term value and stability of scheduling actions. To bound inference complexity and maintain scalability, \schedname evaluates at most \texttt{MAX\_QUEUE\_SIZE = 256} jobs per decision trajectory. The RL agent operates on fixed-size OV and CV embeddings (\(256 \times 8\) and \(256 \times 5\)), applying zero-padding when fewer jobs are present. This design ensures that both the state and action spaces remain constant, allowing the Actor--Critic model stable maintain inference latency across diverse workloads with varying queue lengths. 

Fig.~\ref{fig:Actor-Critic} illustrates the architecture of the Actor–Critic networks along with their inputs and outputs. Both networks are implemented using TensorFlow~\cite{tensorflow}.
The actor network is a three-layer MLP~\cite{MLP} that receives the Observation Vector (OV) as input and outputs an action vector representing job priorities. Each job’s feature set is processed in a sliding-window fashion, enabling the actor to evaluate jobs individually while maintaining global context. After passing the OV through the network and a softmax layer, the resulting normalized priority scores are returned to the environment, which schedules jobs and computes rewards.
The critic network, also a three-layer MLP, processes the Critic Vector (CV). The CV is flattened to include all jobs simultaneously, allowing the critic to estimate the expected cumulative reward for the current job sequence under the actor’s policy. The critic thus provides a scalar value representing the quality of the actor’s decisions.
The actor and critic are trained jointly: after each scheduling decision, the environment returns a reward signal. This reward is used to update the critic’s value estimation and, through backpropagation, to refine the actor’s policy. In this way, the critic guides the actor’s learning, enabling continuous improvement of scheduling decisions across varying workload conditions.

\begin{algorithm}[t!]
\caption{Dynamic Resource Allocation using CVXPY and Mixed-Integer Linear Programming}
\label{algo:11}
\begin{algorithmic}[1]
\Require{cluster\_status, gpus\_per\_node,cpu\_per\_gpu,mem\_per\_gpu}
\Ensure{Decision: way1 (spreading) vs. way2 (packing)}

\State \textcolor{gray}{// Binary variable: 0 for way1, 1 for way2}
\State $x \gets \text{Variable(boolean=True)}$ 
\State $dims \gets (\text{len(cluster\_status)}, \text{gpus\_per\_node})$
\State\textcolor{gray}{// Occupancy matrix}
\State $CJO \gets \text{Variable(dims, boolean=True)}$
\State constraints $\gets []$

\State \textbf{if} way1 is list \textbf{or} way2 is list \textbf{then}
\State \quad \textbf{for all} $(way, val)$ \textbf{in} $[(\text{way1}, 1 - x), (\text{way2}, x)]$ \textbf{do}
\State \quad\quad \textbf{for all} $(node, gpu\_count)$ \textbf{in} way \textbf{do}
\State \quad\quad\quad \textbf{if} node \textbf{in} valid\_nodes \textbf{then}
\State \quad\quad\quad\quad \textbf{for} $g \gets 1$ \textbf{to} gpu\_count \textbf{do}
\State \quad\quad\quad\quad\quad $idx \gets \text{valid\_nodes}[node]$
\State \quad\quad\quad\quad\quad $constraint \gets CJO[idx][g] == val$
\State \quad\quad\quad\quad\quad constraints.append($constraint$)

\State \textbf{for all} $(i, (node\_num, available\_gpus))$ \textbf{in} enumerate(cluster\_status) \textbf{do}
\State \quad \textbf{for} $g \gets 1$ \textbf{to} $\min(\text{available\_gpus}, \text{gpus\_per\_node})$ \textbf{do}
\State \quad\quad total\_occupancy $\gets CJO[i][g]$
\State \quad\quad constraints.append(tot\_occup $\leq$ avail\_gpus)
\State \quad\quad constraints.append($\sum CJO[i] \times$ cpu\_per\_gpu $\leq$ avail\_cpus)
\State \quad\quad constraints.append($\sum CJO[i] \times$ mem\_per\_gpu $\leq$ avail\_mem)
\State \textcolor{gray}{// Maximize GPU occupancy}
\State objective $\gets \text{Maximize(sum(CJO))}$
\State prob $\gets \text{Problem(objective, constraints)}$
\State prob.solve(solver=GLPK\_MI, verbose=False)

\State \textbf{if} $x.value < 0.5$ \textbf{then}
\State \quad selected\_way $\gets$ way1
\State \textbf{else}
\State \quad selected\_way $\gets$ way2

\Return selected\_way
\end{algorithmic}
\vspace{-0.15cm}
\end{algorithm}

\paragraph{\textbf{Reward Function}} The "Score", a job-level reward is calculated based on the targeted optimization goal. For example, if the goal is to minimize wait time, the score is the job’s wait time (scheduled – submitted), and the reward is defined as reward = $-wait\_time$. So the RL agent maximizes reward by reducing wait time. In our design, individual job scores are aggregated because metrics such as average waiting time or bounded slowdown can only be computed after all jobs in a batch are scheduled. Hence, job-level scores are not directly fed to the RL agent. Once the final action in the batch is produced, we compute the performance gap between the baseline (without \schedname) and \schedname’s scheduling outcomes. The normalized difference serves as the reward signal. Feeding rewards as normalized performance gaps reduces variance from sudden fluctuations in input features~\cite{schedinspector, RLScheduler}. This formulation also prevents overfitting to easy scenarios as discussed in Section~\ref{sec:waiting_time_patterns}, ensuring that improvements reflect genuine scheduling gains even under skewed workload patterns and congested queue conditions.      

\paragraph{\textbf{Allocation Optimization Module}} 
The Dynamic Resource Allocation component optimizes job-to-node assignments by selecting the best allocation strategy in real time. When multiple placement options exist (e.g., pack, spread, or hierarchical), this module evaluates the impact of each choice on future cluster performance. As discussed in Sec.~\ref{sec:motivation}, spreading and packing represent key trade-offs between utilization and contention.
Conventional schedulers such as Slurm rely on static heuristics and cannot adapt between these strategies under changing conditions. To overcome this limitation, we model job allocation as a Mixed-Integer Linear Program (MILP) implemented using the GLPK\_MI solver~\cite{glpk} within the CVXPY framework~\cite{cvxpy}. In each iteration, the RL agent outputs a real-time job priority vector. The top-K jobs are passed to the MILP-based optimizer, which selects the optimal job-to-node mapping under current GPU, CPU, and memory constraints. Using look-ahead strategies, it dynamically chooses between spreading and packing to improve long-term cluster performance. 

Algorithm~\ref{algo:11} defines a binary variable \(x\) selects between \textit{way1} (spreading) and \textit{way2} (packing). The occupancy matrix \(CJO\) represents GPU usage per node, with constraints enforcing GPU, CPU, and memory limits. The objective maximizes total GPU occupancy while satisfying per-node resource constraints. After solving, the value of \(x\) determines the selected strategy. As MILP operates on the top-\(K\) jobs prioritized by the RL agent, the formulation considers both current and future job requirements: the top-\(K\) high-priority upcoming jobs in the queue are monitored to explicitly model their resource constraints, start times, and potential usage across multiple time slots.
K is a tunable parameter that can be adjusted according to job burst intensity. 
The Feature Sampling module serves as a bridge between the RL and MILP components, aligning their decisions to complement each other. Feature sampling ensures coordination by emphasizing cues such as future availability and jobs with multiple allocation options. This design allows the RL agent to promote jobs where the optimizer can be most effective, enabling both components to work in tandem toward higher utilization and lower waiting times.

\begin{table*}
\centering
\caption{Trace Summary.}
\vspace{-1em}
\label{table:2}
\normalsize
\resizebox{0.9\textwidth}{!}{ 
\begin{tabular}{|c|c|c|c|c|c|c|c|c|c|c|}
\hline
\textbf{Trace Name} & \textbf{Time} & \textbf{Total Jobs} & \textbf{\makecell{GPUs, \\Number of Nodes}} & \textbf{GPU type} & \textbf{Users} & \textbf{\makecell{Run time\\(avg, max)}} & \textbf{Scheduler} & \textbf{\makecell{Scheduling\\Algorithm}} & \textbf{\makecell{Network\\(same-rack)}} & \textbf{\makecell{Network\\(cross-rack)}} \\ 
\hline
\textbf{Philly'17} & \makecell{Oct’17-Dec’17\\(75 days)} & 96260 & \makecell{2490,\\552} & \makecell{P100(2-GPU),\\P100(8-GPU)} & 319 & \makecell{28329 sec,\\60 days} & \makecell{Apache YARN \\(FIFO)} & Locality-aware & 100-Gbps(Infini-Band) & Ethernet \\
\hline
\textbf{Alibaba'20}  & \makecell{July’20-Aug’20\\(60 days)} & 1.2 million & \makecell{6.5K,\\1.8K} & \makecell{T4, \\Misc,\\ P100, \\V100(16), \\V100(32)} & 1242 & \makecell{4456 sec, \\30 days} &  \makecell{Fuxi \\(FIFO)} & \makecell{GPU Sharing, \\reserving-packing} & \makecell{V100/V100M32: NVlink, \\rest all: PCIe} & Not Allowed\\ 
\hline
\textbf{Helios'21}  & \makecell{April’20-Sep’20} & 1,753K & \makecell{2096,\\262} & \makecell{P100(8-GPU), \\V100(8-GPU)} & 277 & \makecell{6652 sec,\\ 50 days} & \makecell{Slurm \\(FIFO)} & Quasi-Shortest-Service-First & \makecell{Intra-Node:PCIe(Pascal), \\ NVLink(Volta); \\Inter-node: Infiniband} & Not Allowed  \\
\hline
\end{tabular}}
\vspace{-0.7em}
\end{table*}


\section{Experimental Setup}
\label{sec:exp_set}
In this section, we describe the simulated environment, workload traces, scheduling policies, and performance metrics used in our experiments and analysis.
\subsection{Scheduling Environment}
Reinforcement learning requires large amounts of data and interaction, making real-time training impractical. We therefore use a trace-driven simulated environment that enables repeated iterations.
We implemented this environment using OpenAI Spinning Up~\cite{spinningup} and adapted it from RLScheduler~\cite{RLScheduler}, extensively modifying it to support heterogeneous GPUs and fundamentally redefine the learning problem to align with our resource allocation and prioritization logic. The environment loads a job trace and simulates scheduling from an idle cluster. Whenever a job arrives or completes, it selects the next action using our proposed mechanisms. If resources are insufficient, it applies backfilling by placing smaller jobs without delaying higher-priority ones or disrupting running jobs. During training, \schedname{} uses ground-truth runtimes from traces, consistent with prior RL schedulers~\cite{helios,RLScheduler}, to provide accurate reward signals and ensure stable learning. During evaluation, only user-provided (potentially noisy) runtime estimates are used, reflecting realistic conditions. Despite this uncertainty, \schedname{} performs robustly across all baselines. Improved runtime predictors could further enhance overall performance but are orthogonal to the core profiling-free scheduling contributions of this work. Evaluation does not support elastic resource modification (GPU type or count) or preemption within the scope of this work.


\subsection{Workloads/Traces} 
\label{sub:traces}
We evaluate \schedname{} on heterogeneous clusters using real production deep learning (DL) traces: Philly~\cite{philly}, Helios~\cite{helios}, and Alibaba PAI~\cite{MLaaS}. Unlike prior approaches~\cite{gavel, pollux, sia} that replay a small subset of pre-profiled jobs, we use the full traces to drive simulation. Although systems such as~\cite{gavel, pollux, sia} are strong baselines, direct comparison is difficult since their evaluations rely on mapping trace entries to pre-profiled workloads. Their mapped trace version exclude job identifiers and retain only limited metadata (timestamps, durations, GPU counts, GPU time), making it infeasible to map them back to the original traces.
During RL training, each trace is simulated on a representative cluster slice chosen to maintain realistic contention based on job arrivals and runtimes. For example, in Helios, we model five virtual clusters (VC1–VC5) with 16, 12, 10, 8, and 8 nodes (each with 8 GPUs), following prior work~\cite{allox, gandiva, sia}. Table~\ref{table:2} summarizes key trace characteristics, highlighting workload heterogeneity and diversity.
Evaluating on a single trace limits generality~\cite{amvrosiadis2018diversity, patel2020job}. Earlier frameworks (e.g.,~\cite{themis, tiresias, gavel, antman, hived}) relied solely on Philly, then the only public dataset. In contrast, we employ multiple diverse traces, producing policies that generalize across heterogeneous clusters.

\subsection{Scheduling Policies}
\begin{table}
\centering
\caption{Scheduling Policies. Characteristics: submit time (ST), Requested or run time (RT), wait time (WT), Requested Resources (N), and throughput (T).}
\vspace{-1em}
\label{table:table_policies}
\resizebox{\columnwidth}{!}{ 
\begin{tabular}{|l|r|r|}
\hline
\textbf{Scheduling Policy} & \textbf{Characteristics} & \textbf{Equation or Strategy} \\ 
\hline
\textbf{FCFS}   & ST   & St\\
\hline
\textbf{SJF}  & RT    &  Rt\\ 
\hline
\textbf{WFP3\cite{tang2009fault}}  & ST, RT, WT, N &  $-\left(\frac{\text{wt}}{\text{rt}}\right)^3 \times \text{nt}$  \\
\hline
\textbf{UNICEP\cite{tang2009fault}}  & ST, RT, WT, N &  $-\frac{\text{wt}}{\log_2(\text{nt}) \times \text{rt}}$  \\
\hline
\textbf{F1\cite{carastan2017obtaining}}  & ST, RT, N &  $\log_{10}(\text{rt}) \times \text{nt} + 870 \times \log_{10}(\text{st})$  \\
\hline
\end{tabular}
}
\vspace{-1em}
\end{table}

We compare \schedname{} against several scheduling policies, including FIFO, SJF, WFP3, UNICEP~\cite{tang2009fault}, F1~\cite{carastan2017obtaining}, Slurm Multifactor Priority. These policies rely on different job characteristics, summarized in Table~\ref{table:table_policies}. FIFO schedules jobs by submission order, while SJF favors shorter runtimes. WFP3 and UNICEP~\cite{tang2009fault} combine multiple factors, prioritizing jobs with short runtimes, small resource requests, or long waits, often tuned with expert knowledge. The F1 scheduler~\cite{carastan2017obtaining} is ML-based, using brute-force simulation and non-linear regression to minimize target metrics.

\subsection{Performance Metrics}
We evaluate four key performance metrics: (1)\textbf{Wait Time}: The time between a job's submission and its start. (2) \textbf{Job Completion Time (JCT)}: The mean duration from submission to completion, equal to wait time plus runtime averaged across all jobs. (3)  \textbf{Average bounded slowdown (bsld)}: Measures slowdown relative to runtime, balancing penalties for both long-waiting and long-running jobs. Introduced in~\cite{BSLD1998}. (4)  \textbf{Resource utilization}:  The mean percentage of allocated GPUs, normalized by the total GPUs in the cluster over time.

\begin{figure}[!ht]
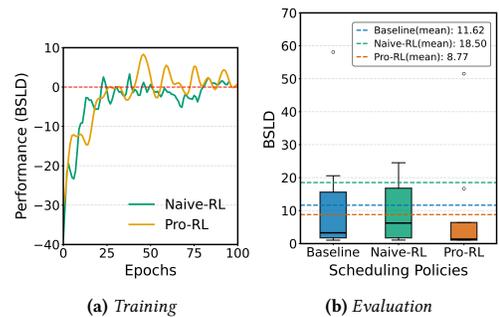

\centering
\begin{subfigure}[b]{0.18\textwidth}
    \centering
    \includegraphics[width=\textwidth]{figures/Eval_1/train_pro.pdf}
    \caption{Training}
    \label{subfig:training_naiveproRL}
\end{subfigure}
\begin{subfigure}[b]{0.18\textwidth}
    \centering
    \includegraphics[width=\textwidth]{figures/Eval_1/pro_new.pdf}
    \caption{Evaluation}
    \label{subfig:eval_naiveprorl}
\end{subfigure}
\vspace{-1em}
\caption{Comparison between performance of naive-RL and pro-RL}
\vspace{-0.5em}
\label{fig:traineval_naiveproRL}
\vspace{-0.1cm}
\end{figure}

\begin{figure*}[!ht]
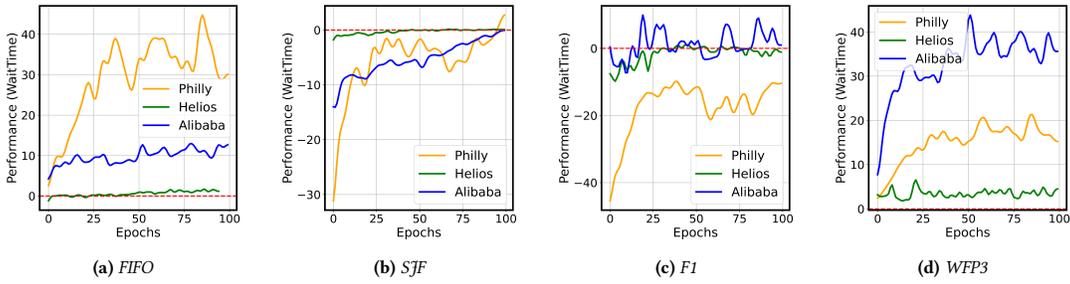

    \centering
    \begin{subfigure}{0.18\textwidth}
        \includegraphics[width=\textwidth]{figures/Eval_1/11_1.pdf}
        \caption{FIFO}
    \end{subfigure}
    \hspace{1em}
    \begin{subfigure}{0.18\textwidth}
        \includegraphics[width=\textwidth]{figures/Eval_1/11_2.pdf}
        \caption{SJF}
    \end{subfigure}
    \hspace{1em}
    \begin{subfigure}{0.18\textwidth}
        \includegraphics[width=\textwidth]{figures/Eval_1/11_3.pdf}
        \caption{F1}
    \end{subfigure}
    \hspace{1em}
    \begin{subfigure}{0.18\textwidth}
        \includegraphics[width=\textwidth]{figures/Eval_1/11_4.pdf}
        \caption{WFP3}
    \end{subfigure}
    \vspace{-1em}
    \caption{Training curves of \schedname on three real-world traces for the performance metric \textit{Wait Time}, using four different base policies. Subfigures highlight the respective base policy, and the y-axis shows the performance difference between \schedname and the corresponding base policy.}
    \vspace{-0.5em}
    \label{fig:1_train}
\end{figure*}

\section{Evaluation}
\label{sec:eval}
We present evaluation of \schedname under various scenarios. 

\vspace{-0.35em}
\subsection{Design Choice Justification.} This section evaluates the key design choices in \schedname, focusing on the impact of feature construction, sampling strategy, and CVXPY-based allocation. We first analyze how these components affect learning efficiency and overall scheduling performance. 

\noindent\textbf{Naive-\schedname vs Pro-\schedname}. We evaluate two variants of \schedname: naive-\schedname and pro-\schedname. In naive-\schedname, raw trace features are fed directly to the RL agent without feature construction or allocation optimization. Since MILP-based allocation is disabled, job placement follows Slurm’s default (OverSubscribe=No). Pro-\schedname adds two enhancements, feature sampling and MILP-based allocation optimization. The engineered features capture key aspects of cluster state, job queue, and job characteristics, as described in~\cref{subsec:sys_comp}, while the solver ensures resource-aware placement.
Figure~\ref{subfig:training_naiveproRL} shows their training curves using Slurm as a baseline on the Philly trace, and Figure~\ref{subfig:eval_naiveprorl} presents the evaluation results. Pro-\schedname achieves a 52.59\% improvement in BSLD over naive-\schedname. This result highlights that directly feeding raw features to RL and expecting it to learn everything performs poorly in complex GPU environments. However, structured feature design and solver-guided allocation significantly enhance learning efficiency and scheduling performance.
In all subsequent experiments, we refer to pro-\schedname simply as \schedname.

\subsection{Generalization Across Diverse Job Traces and Base Scheduling Policies}

\begin{figure*}[!ht]
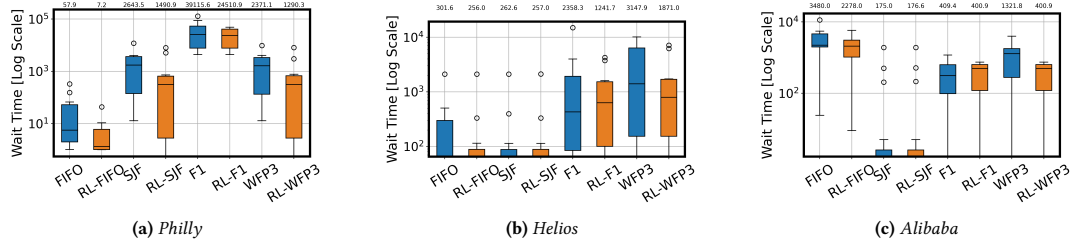

    \centering
    \begin{subfigure}[b]{0.24\textwidth}
        \centering
        \includegraphics[width=\textwidth]{figures/Eval_1/12_1.pdf}
        \vspace{-1.5em}
        \caption{Philly}
        \label{fig:eval_philly}
    \end{subfigure}
    \hspace{2em}
    \begin{subfigure}[b]{0.24\textwidth}
        \centering
        \includegraphics[width=\textwidth]{figures/Eval_1/12_2.pdf}
        \vspace{-1.5em}
        \caption{Helios}
        \label{fig:eval_helios}
    \end{subfigure}
    \hspace{2em}
    \begin{subfigure}[b]{0.24\textwidth}
        \centering
        \includegraphics[width=\textwidth]{figures/Eval_1/12_3.pdf}
        \vspace{-1.5em}
        \caption{Alibaba}
        \label{fig:eval_alibaba}
    \end{subfigure}
    \vspace{-1em}
    \caption{Waiting time distribution across base policies and RL policy for three traces. Mean values of the performance metric are annotated on top of each graph.}
    \label{fig:combined_eval}
\end{figure*}

We evaluate \schedname on three heterogeneous traces, Philly, Helios, and Alibaba, each representing a distinct cluster configuration with unique workload characteristics (Tables~\ref{tab:cpu_gpu},~\ref{table:2}). This diversity ensures that \schedname is tested across varied workload patterns and cluster conditions.
In each training epoch, the RL agent processes 100 batches of 256 jobs, updating the actor–critic policy and value networks after every batch. The agent is trained to enhance performance over four base scheduling policies: FIFO, SJF, F1, and WFP3.
Figure~\ref{fig:1_train} shows the training curves for 100 epochs using wait time as the performance metric. Each curve corresponds to a specific policy, with results from the three traces normalized to the range 
$-50$ to $+50$ for consistent scaling and compact visualization.

Figure~\ref{fig:1_train} illustrates the training progress of \schedname, showing the normalized difference between the RL agent’s performance and each corresponding base scheduling policy (FIFO, SJF, F1, and WFP3) for the wait time metric. A rising curve indicates that the RL agent is learning to reduce average wait time more effectively than the baseline policy. The eventual flattening of the curve signifies that training reaches convergence and the policy stabilizes. Across traces, the vertical scale differs because of workload diversity. 


After training, we evaluate \schedname’s performance against the four base scheduling policies as shown in Fig.~\ref{fig:combined_eval}. For each workload trace, jobs are scheduled using both the original policy and its RL-enabled counterpart (e.g., FIFO vs. RL-FIFO). Each experiment runs ten times with random sequences of 1,024 jobs, and the average wait time is reported.
Overall, \schedname consistently lowers average wait time relative to its base policies. For instance, in the Philly trace under FIFO, RL-FIFO achieves an 87.5\% reduction in wait time, demonstrating a substantial gain in scheduling efficiency. In the Alibaba trace under SJF, the results show only a small change, reflecting limited room for improvement in a workload already suited to SJF. These results confirm that \schedname effectively adapts across diverse policies and traces, enhancing multiple scheduling strategies under a unified framework.

\begin{figure}[!ht]
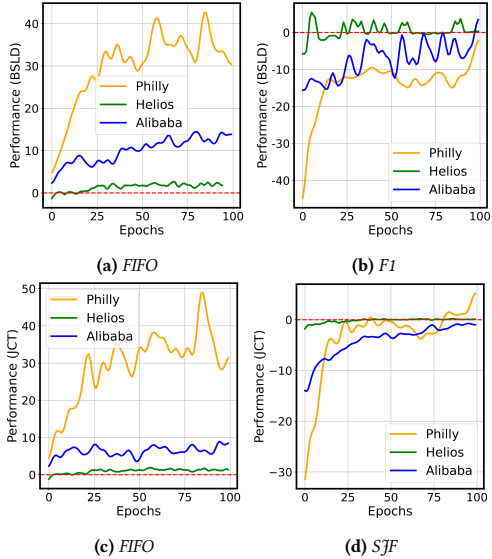

    \centering
    \begin{subfigure}{0.18\textwidth}
        \centering
        \includegraphics[width=\textwidth]{figures/Eval_2/13_1.pdf}
        \caption{FIFO}
        \label{fig:fifo_bsld}
    \end{subfigure}
    \begin{subfigure}{0.18\textwidth}
        \centering
        \includegraphics[width=\textwidth]{figures/Eval_2/13_2.pdf}
        \caption{F1}
        \label{fig:f1_bsld}
    \end{subfigure}
    \begin{subfigure}{0.18\textwidth}
        \centering
        \includegraphics[width=\textwidth]{figures/Eval_2/13_3.pdf}
        \caption{FIFO}
        \label{fig:fifo_jct}
    \end{subfigure}
    \begin{subfigure}{0.18\textwidth}
        \centering
        \includegraphics[width=\textwidth]{figures/Eval_2/13_4.pdf}
        \caption{SJF}
        \label{fig:sjf_jct}
    \end{subfigure}
    \vspace{-1em}
    \caption{Training curves of \schedname for BSLD and JCT across three traces, trained using two base scheduling policies.}
    \vspace{-.5em}
    \label{fig:one_by_six}
\end{figure}

\subsection{Performance on Multiple Metrics}
This section evaluates \schedname across several metrics, bounded slowdown (BSLD), job completion time (JCT), and GPU utilization. While waiting-time results were discussed earlier, here we focus on the remaining metrics and compare \schedname against three base scheduling policies. Results for additional policies are omitted due to space limits. The training and evaluation setup remains identical to the previous section. 
Figures \ref{fig:fifo_bsld} and \ref{fig:f1_bsld} show training curves of \schedname when trained with FIFO and F1 as base policies for the BSLD metric. Figures \ref{fig:fifo_jct} and \ref{fig:sjf_jct} present the corresponding results of \schedname for JCT under FIFO and SJF as base policies. 

\begin{figure}[!ht]
    \centering
    \includegraphics[width=0.3\textwidth]{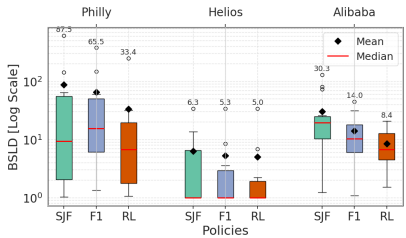} 
    \vspace{-1em}
    \caption{BSLD distribution across different base policies and the RL policy for three traces. Lower values indicate better performance.}
    \vspace{-1em}
    \label{fig:2_eval_BSLD}
\end{figure}

\begin{figure}[!ht]
    \centering
    \includegraphics[width=0.3\textwidth]{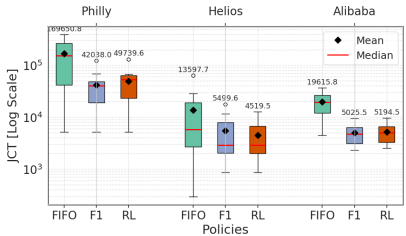} 
    \vspace{-1em}
    \caption{Job Completion Time distribution across different base policies and the RL policy for three traces. Lower values indicate better performance.}
    \vspace{-1em}
    \label{fig:2_eval_JCT}
\end{figure}


\begin{table}[!ht]
    \centering
    \caption{Utilization improvement across policies and traces}
    \vspace{-0.7em} 
    \label{tab:uti_table}
    \renewcommand{\arraystretch}{0.85} 
    \begin{tabular}{l S S S}
        \toprule
        & \textbf{FIFO} & \textbf{SJF} & \textbf{F1} \\
        \midrule
        \textbf{Philly}  & 4.83\% & 3.44\% & 13.62\% \\
        \textbf{Helios}  & 1.00\% & 19.71\% & 1.95\% \\
        \textbf{Alibaba} & 2.40\% & 11.30\% & 1.28\% \\
        \bottomrule
    \end{tabular}
    \vspace{-0.6em}
\end{table}

Figures \ref{fig:2_eval_BSLD} and \ref{fig:2_eval_JCT} present evaluation results of \schedname compared to base scheduling policies for the BSLD and JCT metrics across the three traces. As shown in Table~\ref{tab:uti_table}, \schedname improves cluster utilization across all workloads relative to the baseline policies. Across workloads, \schedname reduces BSLD by at least 5.28\% over F1 on Helios and up to 72.32\% over SJF on Alibaba. For JCT, it achieves up to 70.68\% improvement over FIFO on Philly, with a modest 1.3\% degradation relative to F1 on Alibaba, likely because F1 already attains near-optimal job placements in that trace. \schedname increases resource utilization by 13.62\% and 19.71\% on Philly and Helios, respectively, relative to the F1 and SJF policies.

The observed trends align with workload characteristics and the nature of baseline policies. Philly’s mix of long, multi-GPU jobs allows \schedname to learn more effective prioritization, yielding substantial gains in wait time, BSLD, and utilization. Helios and Alibaba, dominated by short jobs, offer limited headroom since heuristics like SJF and F1 already perform near-optimally. While SJF assumes perfect knowledge of job runtimes and F1 relies on static log-scaled features, both lack adaptability to evolving queue states and heterogeneous resources. In contrast, \schedname learns directly from runtime signals, generalizing across traces and achieving consistent improvements in BSLD and utilization even when JCT gains are modest. These results reflect the complementary nature of the evaluated metrics, where wait time and JCT capture per-job responsiveness, BSLD and utilization reveal system-level efficiency. Nevertheless, \schedname underscores its versatility and effectiveness across multiple performance dimensions.

\subsubsection{Transfer Learning}
\label{subsub:cross_test}
\begin{table}[ht]
\centering
\caption{Wait time improvement on the Helios trace using cross-policy models.
\vspace{-0.3cm}
}
\label{table:cross_policy}
\resizebox{0.3\textwidth}{!}{ 
\begin{tabular}{|c|c|c|c|c|}
\hline
\multirow{2}{*}{\textbf{trained on}} & \multicolumn{4}{c|}{\textbf{tested on}} \\ \cline{2-5} 
 & FIFO & SJF & F1 & WFP3 \\ \hline
FIFO &14.95\%  &2.29\%  &4.73\%  &17.32\%  \\ \hline
SJF  &14.22\%  &2.22\%  &4.70\%  &15.23\%  \\ \hline
F1   &14.91\%  &1.66\%  &4.73\%  &13.81\%  \\ \hline
WFP3  &3.76\%  &-4.01\%  &-4.31\%  &11.83\%  \\ \hline
\end{tabular}
}

\vspace{-0.2cm}
\end{table}
Our training setup is not limited to a specific policy or metric; instead, we assess how a model trained under one policy performs when used to make scheduling decisions under another. Table~\ref{table:cross_policy} reports percentage improvements in wait time on the Helios trace when each model is trained on one base policy and tested against all others. The results show that \schedname generalizes effectively across policies within the same trace, reducing the need for frequent retraining. Performance degradation is observed when trained on WFP3 and tested on other policies like SJF, and F1. SJF and F1 prioritize short, early jobs, which sometimes aligns with WFP3’s fairness model. However, WFP3-trained agents emphasize long waiting jobs, even if they can be large or late, leading to mismatches under SJF and F1 (linear and log-scaled score) and may have reduced performance. Thus, training cost can be reduced by leveraging transfer, showing potential to lower the overhead of training in dynamic environments.

\subsection{Slurm in Simulated Environment} 
\begin{figure}[!ht]
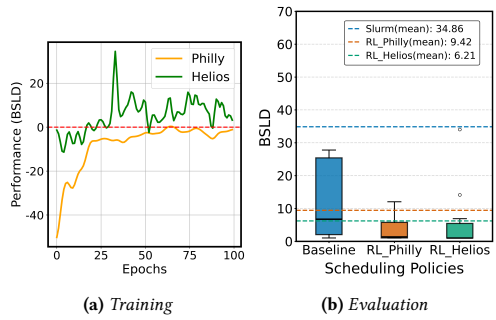

\centering
\begin{subfigure}[b]{0.18\textwidth}
    \centering
    \includegraphics[width=\textwidth]{figures/Motivation/16_train.pdf}
    \caption{Training}
    \label{subfig:training_design}
\end{subfigure}
\begin{subfigure}[b]{0.18\textwidth}
    \centering
    \includegraphics[width=\textwidth]{figures/Eval_2/16_b.pdf}
    \caption{Evaluation}
    \label{subfig:eval_design}
\end{subfigure}
\vspace{-1em}
\caption{Results on Philly and Helios trace using SLurm as a base scheduler.}
\label{fig:design_jus}
\end{figure}

Until now, we have examined the effectiveness and generality of \schedname based on various heuristic scheduling policies. The Slurm multifactor priority plugin adjusts job order using a weighted blend of age\_factor, fair\_share, job\_attributes, and partition\_factor, QoS\_factor. age\_factor is waiting time, fairshare\_factor by mapping CPU‑based fair‑share math to GPU. We use the job’s requested run time as the job\_attribute\_factor, and the partition\_factor represents the priority assigned to each queue in the system. In our study, we approximate each factor for GPUs and set all weights to 1000 to ensure equal contribution. Figure~\ref{fig:design_jus} presents training curves on the Helios and Philly traces, comparing bounded slowdown (BSLD) against Slurm’s baseline. \schedname consistently outperforms Slurm, reducing BSLD by 71.54\% on Philly and 81.18\% on Helios, demonstrating that it can surpass Slurm’s multifactor priority mechanism in simulation.

\subsection{Comparison against State-of-the-Art Systems}
\label{sub:SOTA}

Finally, we compare \schedname with SOTA scheduler Quasi-Shortest-Service-First Scheduler (QSSF)~\cite{helios}. The QSSF
policy uses history-based job priority predictions, and the authors have made these predictions publicly available for the Philly trace. Hence we can compare our system against QSSF scheduler for Philly trace. The results are presented in Table~\ref{table:qssf}. 
For all four performance metrics, we see significant improvement with \schedname as compared to QSSF.  \schedname brings 25\% improvement in wait time 
3.25$\times$ better performance on the BSLD metric.

To verify the robustness of \schedname, we conduct a large-scale experiment using 10,000 consecutively executed jobs, comparing \schedname against QSSF. In this setup, Fig.~\ref{fig:qssf_eval} presents the job completion time (JCT), which better reflects long-term system efficiency. The results show a 48.43\% improvement in JCT with \schedname. As noted in Table~\ref{tab:cpu_gpu}, the Philly trace exhibits high wait and run times, showing the benefits of adaptive scheduling under heavy and long-running workload.

\begin{minipage}{0.23\textwidth}
    \centering
    \vspace{+1em}
    
    \resizebox{\linewidth}{!}{%
    \begin{tabular}{|l|l|l|}
    \hline
    \textbf{Performance} & \textbf{QSSF} & \textbf{\schedname} \\ \hline
    \textbf{Wait Time} & 3748.14 & 2830.01 \\ \hline
    \textbf{BSLD} & 28.17 & 20.11 \\ \hline
    \textbf{JCT} & 35567.97 & 33199.58 \\ \hline
    \textbf{Utilization} & 4.72 & 4.97 \\ \hline
    \end{tabular}%
    }
    \vspace{+0.5em}
    \captionof{table}{Comparison of QSSF and \schedname performance (with backfilling).}
    \label{table:qssf}
\end{minipage}
\hfill
\begin{minipage}{0.22\textwidth}
    \centering
    \includegraphics[width=\linewidth]{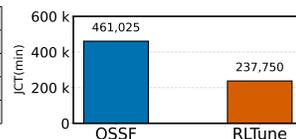}
    \vspace{-2.5em}
    \captionof{figure}{Comparison of JCT for QSSF and \schedname\ on 10k jobs.}
    \label{fig:qssf_eval}
\end{minipage}



\begin{table*}[t]
\centering
\footnotesize
\setlength{\tabcolsep}{4pt}
\renewcommand{\arraystretch}{1.2}
\caption{Comparison of scheduling policies across three traces: Philly, Helios, and Alibaba. WT (s) = Wait Time, JCT (s) = Job Completion Time, BSLD = Bounded Slowdown, Utilization (\%) = GPU utilization. “Time” denotes total elapsed time in seconds.}
\vspace{-1.0em}
\begin{tabular}{l|cccc|cccc|cccc|c}
\toprule
\textbf{} & \multicolumn{4}{c|}{\textbf{Philly}} & \multicolumn{4}{c|}{\textbf{Helios}} & \multicolumn{4}{c|}{\textbf{Alibaba}} & \textbf{Time} \\
\textbf{Policy} & BSLD & WT & JCT & Util & BSLD & WT & JCT & Util & BSLD & WT & JCT & Util &  \\
\midrule
FIFO           & 1298.06 & 142391.50 & 169650.84 & 4.63 & 161.77 & 7264.58  & 10483.16 & 4.03 & 88.32 & 14306.14 & 19219.08 & 0.08 & 86.76 \\
SchedInspector & 1114.52 & 120623.63 & 150347.88 & 4.85 & 152.16 & 7187.54  & 10715.42 & 4.00 & 89.52 & 14470.00 & 19382.94 & 0.09 & 96.13 \\
RLScheduler    & 491.92  & 86512.85  & 112022.08 & 4.64 & 117.83 & 6178.47  & 9211.55  & 4.01 & 52.18 & 9641.81  & 14554.75 & 0.06 & 181.27 \\
\schedname         & 232.82  & 34199.93  & 85761.51  & 4.85 & 75.24  & 4335.87  & 7541.40  & 4.09 & 44.05 & 8726.45  & 13639.38 & 0.09 & 102.38 \\
\bottomrule
\end{tabular}
\label{tab:scheduler_comparison}
\end{table*}

Table~\ref{tab:scheduler_comparison} presents a comparative analysis of multiple scheduling policies across three large-scale GPU traces.  We evaluate each policy using five key metrics BSLD, wait time, JCT, GPU utilization, and time which is a scheduling overhead for 10 batches of 256 jobs. The FIFO policy is just for the baseline. We compare \schedname against two most recent state-of-the-art RL-based schedulers: RLScheduler and SchedInspector allowing a direct comparison of performance and execution time. Notably, the original simulated environments used in these papers do not support GPU workloads. To ensure a fair and meaningful comparison, we reimplemented the core RL mechanisms of each scheduler and adapted them to run on GPU traces. 
Across all three traces, \schedname delivers the best overall performance, achieving the lowest BSLD, wait time, and JCT while maintaining high utilization. On Philly, it reduces BSLD to 232.82 and JCT to 85.7k s, outperforming RLScheduler (491.92) and SchedInspector (1114.52). On Helios, it attains a BSLD of 75.24 with 43.3\% lower wait time than RLScheduler, maintaining comparable utilization (4.09) and moderate execution time (102 s). On Alibaba, \schedname achieves BSLD (44.05) and low wait time (8726 s), surpassing all baselines. While RLScheduler shows reasonable performance, its runtime overhead (181 s) is the highest. Overall, \schedname achieves an effective balance between scheduling performance and runtime efficiency across heterogeneous GPU workloads.

\subsection{Real Slurm Deployment}To evaluate end-to-end system performance, we conducted an experiment using a heterogeneous Slurm cluster comprising two P100 nodes (each with 4 GPUs), two K80 nodes (each with 2 GPUs), and one M40 node (with 1 GPU). We used \textit{Slurm v21.08.5} configured with \textit{SchedulerType=sched/backfill}, \textit{SelectType=select/cons\_tres}, and \textit{PriorityType=priority/multifactor} with equal weights assigned to all priority factors. 
We generated a synthetic trace of 1,024 ML/DL jobs on a heterogeneous Slurm cluster, consisting primarily of deep learning fine-tuning and inference workloads. These included large language model (LLM) jobs requiring a single GPU, multiple GPUs, or multi-node execution. The dynamically arriving batch was submitted to both Slurm’s weight-tuned Multi-factor Priority scheduler and the \schedname-equipped scheduler under identical conditions. \schedname adjusted job priorities at 1-minute intervals and applied a custom allocation mechanism that dynamically toggled the \textit{OverSubscribe=NO} flag. 


\begin{figure}[!ht]
    \centering
    \begin{minipage}{0.23\textwidth}
        \centering
        \includegraphics[width=\linewidth]{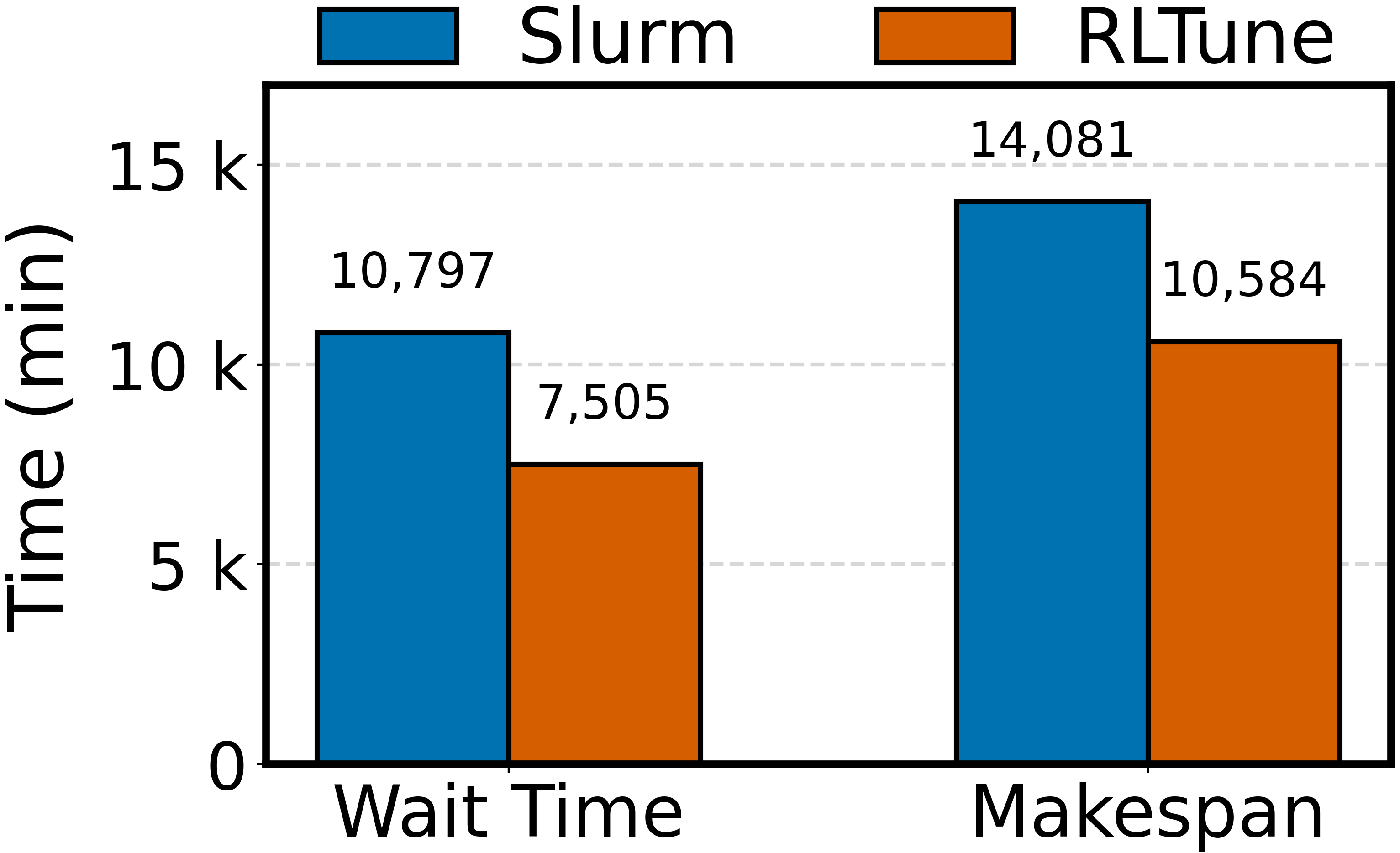}
        \vspace{-1em}
        \caption{End-to-end performance of \schedname vs slurm.}
        \label{fig:endtoend}
    \end{minipage}
    \vspace{-1em}
    \hfill
    \begin{minipage}{0.22\textwidth}
        \centering
        \includegraphics[width=\linewidth]{figures/eval_realslurm/end_uti.pdf}
        \vspace{-1em}
        \caption{End-to-end performance: utilization \schedname vs slurm.}
        \label{fig:end_moti}
        
    \end{minipage}
\end{figure}

The results, summarized in Fig.~\ref{fig:endtoend},~\ref{fig:end_moti}, show that \schedname{} achieved a makespan of 10,584~mins compared to 14,081~mins for the Multi-factor Priority Slurm baseline (a 24.8\% reduction), increased overall GPU utilization by 3.9\%, and reduced average wait time by 30.5\%.



\subsection{Operation Costs}
\label{sec:operation_costs}


Training \schedname{} takes between 3--8 hours, depending on the trace length and job diversity. During inference, the system to maintain inference latency of \(\sim0.7~\text{ms}\) (including state construction and RL forward pass), and the \texttt{CVXPY} solver adds only \(\sim0.2~\text{ms}\), which is acceptable for batch-scheduling workloads. 
To evaluate scalability, we stress-tested \schedname{} with up to 10{,}000 concurrent job arrivals. Decision latency increases sub-linearly from 7.8~s, 9.8~s, 14.3~s, and 22.8~s for queue sizes of 128, 256, 512, and 1024, respectively corresponding to 1.26\(\times\)--1.59\(\times\) growth as the queue doubles. 
The MILP solver is triggered only when multiple placements exist for high-priority jobs (typically \(H = 8\text{--}16\)). Under bursty arrivals, this parameter can be reduced to meet latency budgets.


\section{Related Work}
\label{sec:related-work}

\noindent\textbf{Scheduling in Heterogeneous Environments.}
Scheduling in heterogeneous GPU clusters has been approached in various ways.
Systems such as Gavel \cite{gavel} and Pollux \cite{pollux} optimize throughput by leveraging job-level profiling to predict performance under different resource allocations, while Sia \cite{sia} extends this approach to heterogeneous GPUs through co-adaptive tuning. However, their dependence on pre-profiled workloads and application-specific metadata 
limits adaptability and fair trace comparison.
In contrast, \schedname learns scheduling policies dynamically from real-time job and cluster states, generalizing to unseen workloads without profiling overhead.
Other schedulers address orthogonal objectives.
Lucid~\cite{lucid} focuses on scheduling interpretability, and
Lyra~\cite{lyra} explores elastic scheduling by loaning idle inference GPU servers for elastic training jobs.
Shockwave~\cite{shockwave} targets job progress fairness using stochastic dynamic programming, explicitly handling temporal variations in job throughput.
Distributed LLM serving systems like FairServe~\cite{khan2024fairserve} and training schedulers like like Optimus~\cite{optimus}, Tiresias~\cite{tiresias}, and Gandiva~\cite{gandiva} aim to improve JCT, efficiency, and fairness on heterogeneous setups.
Others further consider JCT fairness for training jobs~\cite{themis,gandivafair,hived}, 
Antman~\cite{antman} for co-location on homogeneous GPUs, while Allox~\cite{allox} for CPU-GPU interchangeability.
SchedTune~\cite{schedtune} incorporates ML-based predictions using historical data.
Recent efforts expand to network-, storage-, and elasticity-aware scheduling.
CASINNI~\cite{cassini} mitigates communication bottlenecks,
Easyscale~\cite{easyscale} supports elastic training,
FDG~\cite{Alibaba23} reduces GPU fragmentation,
and SiloD~\cite{silod}, SHADE~\cite{khan2023shade}, and FedCaSe~\cite{khan2024fedcase} integrate caching with scheduling.
Acme~\cite{acme} characterizes large-scale LLM workloads, motivating adaptive and fault-tolerant scheduling. 

Existing schedulers rely on profiling, heuristics, or narrow objectives. \schedname combines RL-based prioritization and MILP-based allocation for profiling-free scheduling across heterogeneous GPUs.

\noindent\textbf{Reinforcement Learning for Task Scheduling.} While RLScheduler~\cite{RLScheduler}, SchedInspector~\cite{schedinspector}, MARS~\cite{mars}, ~\cite{li2023batch}, and DRAS~\cite{DRAS} advance RL-based scheduling for CPU workloads, they operate under static or workflow-oriented models. Their learning focuses on CPU-centric environments. In contrast, \schedname{} targets GPU-intensive ML/DL workloads to extend this line of work toward GPU-aware, multi-resource scheduling, where learning-based prioritization and solver-based optimization shows potential to address fine-grained resource contention and dynamic workload diversity.

\section{Conclusion}
\label{sec:conclusion}

In this work, we addressed the challenge of scheduling ML/DL workloads on large GPU clusters through \schedname, an application-agnostic RL+MILP-based dynamic scheduling policy. Unlike prediction-based schedulers, \schedname jointly optimizes prioritization and allocation without per-application profiling, adapting to dynamic workloads.
Evaluations on real-world traces show up to 81\% lower queueing delay, 70.8\% shorter JCT, and 20\% higher GPU utilization, along with runtime cost savings. Compared with state-of-the-art RL schedulers, \schedname achieves 1.2× faster job completions, up to 35\% lower waiting times, and 20\% higher utilization across diverse traces. On a heterogeneous Slurm cluster with a 1,024-job synthetic trace, it further reduces makespan by 24.8\%, improves utilization by 3.9\%, and lowers wait time by 30.5\%.
Overall, \schedname generalizes across cluster types and objectives, providing a robust and adaptive solution for heterogeneous GPU scheduling. RLTune Github Link ~\cite{rltune_github}.

\begin{acks}
We thank the anonymous reviewers for their valuable feedback. Some results were obtained using the Chameleon testbed~\cite{chameleon_cloud}, supported by the NSF. This work is supported in part by NSF grants CSR-2106634, CSR-2312785, National Natural Science Foundation of China, Grant no. 62202382.
\end{acks}

\bibliographystyle{ACM-Reference-Format}
\bibliography{sample-base}

\end{document}